\documentclass[epj,final]{svjour}
\usepackage{graphics,epsfig,rotating}

\begin{document}

\title{
What is the physics behind the $^3$He--$^4$He anomaly ?
}

\author{
     W. Neubert\inst{1}
and  A.S. Botvina\inst{2}
}
\institute{
     Institut f\"ur Kern- und Hadronenphysik, Forschungszentrum Rossendorf, 
     01314 Dresden, Germany 
\and Istituto Nazionale di Fisica Nucleare, 40126 Bologna, Italy \\
     {\em and} Institute for Nuclear Research, Russian Academy of Science,
     117312 Moscow, Russia
}

\date{Received: \today}

\abstract{We show that coalescence of nucleons emitted prior to 
thermalization in
highly excited nuclei can explain the anomaly of kinetic energies of 
helium clusters. A new coalescence algorithm has been included in
the statistical approach to 
nuclear reactions formerly used to describe intermediate mass fragment 
production.}
 
\maketitle

\section{Introduction}

In this paper we address a phenomenon which seems to be important 
for understanding mechanisms which favour light cluster production in  
intermediate energy reactions . It concerns the so-called 
$^3$He--$^4$He "puzzle", i.e.
the anomalous behaviour of their kinetic energies.
 Usually, thermal models (e.g. 
\cite{mekjian}) or approaches which consider a possible radial expansion 
(flow) \cite{siemens} 
are applied to describe the kinetic energies of light charged particles 
(LCP). In the thermal scenario we expect 
$\langle E_{kin}(^3He) \rangle \approx \langle E_{kin}(^4He) \rangle$ whereas 
radial flow delivers 
$\langle E_{kin}(^3He) \rangle < \langle E_{kin}(^4He) \rangle$.  
However, in many reactions quite an opposite behaviour has been observed. 
The corresponding data are summarized in table 1.
For example, there is evidence of the "puzzle" in the inclusive kinetic energy 
spectra of He isotopes obtained from p + Ag \cite{Vol74} and
p+C \cite{And98} reactions at 1 GeV, in 7.5 GeV/c proton collisions with
$^{12}$C, $^{112,124}$Sn,$^{197}$Au \cite{Bog80} and in Ne+U reactions at
250 and 400 A$\cdot$MeV \cite{gutbrod}. The anomaly was also observed
in antiproton reactions 202 MeV/c $\bar p$ + 
$^{12}$C,$^{40}$Ca,$^{63}$Cu,$^{92,98}$Mo and $^{238}$U \cite{Mar88}.  
Precise measurements of 55 MeV $^3$He+Ag collisions at $\Theta_{lab}$=147.5$^o$
also have shown the anomalous behaviour of $^3$He and $^4$He 
(ref.\cite{Vio98}). The following trends have been observed in heavy-ion 
collisions: 
(i) the difference  $\Delta$E=
$\langle E_{kin}(^3He) \rangle$--$\langle E_{kin}(^4He) \rangle$
is positive and increases with the particle multiplicity in
Au + Au reactions at 250 A$\cdot$MeV \cite{Dos87},
(ii) the "puzzle" is pronounced in central event samples of
Au + Au \cite{Pog93} , 95 A$\cdot$MeV Ar + Ni \cite{Bor96},
\mbox{50 A$\cdot$MeV} Xe+Sn \cite{Bou99} but weaker in
Ar + Ca \cite{Moh96} collisions, 
(iii) positive values $\Delta$E are observed in the incident 
energy range from about 50 A$\cdot$MeV to 300 A$\cdot$MeV but the anomaly
vanishes at $\simeq$ 1 GeV \cite{Lis95}. The existing data indicate that
the larger the number of nucleons of the colliding system the larger the
deviation of the kinetic energies of $^3$He and $^4$He. 
 
\begin{table}
\caption{Experimental data concerning the $^3$He-$^4$He
anomaly. The temperatures T of the helium isotopes were obtained by Maxwell-Boltzmann
fits to the kinetic energy distributions, in case of $^*$)
an exponential fit to the data at $\Theta_{lab}$=90$^o$ was applied.}
\label{tab:data}
\begin{center}
\begin{tabular}{|l|c|c|c|}
\hline\noalign{\smallskip}
Reaction&Incident energy &   T($^3$He)     &  T($^4$He)              \\ 
        &  or momentum   &   (MeV)         &  (MeV)                  \\    
\hline
p+$^{12}$C~\cite{And98}    &  1 GeV     &  6.8$\pm$0.2 &  5.8$\pm$0.2  \\
\hline
p+Ag~\cite{Vol74}           &   1 GeV    & 11.2$\pm$0.2 &  4.6$\pm$0.1  \\   
\hline
$\bar p+^{12}$C~\cite{Mar88}& 202 MeV/c  & 19.8$\pm$12. &  15.6$\pm$9.  \\
\hline    
$\bar p+^{40}$Ca~\cite{Mar88}& 202 MeV/c & 24.2$\pm$13. &  16.8$\pm$9   \\
\hline 
$\bar p+^{63}$Cu~\cite{Mar88}& 202 MeV/c & 22.2$\pm$10. &  15.4$\pm$8.  \\
\hline
$\bar p+^{92}$Mo~\cite{Mar88}& 202 MeV/c & 25.2$\pm$17. &  16.2$\pm$9.   \\
\hline
$\bar p+^{98}$Mo~\cite{Mar88}& 202 MeV/c & 21.8$\pm$6.  &  17.1$\pm$7.   \\
\hline
$\bar p+^{238}$U~\cite{Mar88}& 202 MeV/c & 20.1$\pm$12. &  14.2$\pm$4.   \\          
\hline
p+$^{12}$C~\cite{Bog80} &  7.5 GeV/c     & 37.$\pm$2.   &  28.$\pm$2.    \\ 
\hline
p+$^{112}$Sn~\cite{Bog80} &  7.5 GeV/c    & 38.$\pm$1.   &  27.$\pm$1.  \\          
\hline
p+$^{124}$Sn~ \cite{Bog80}&  7.5 GeV/c    & 43.$\pm$1.   &  27.$\pm$1.   \\
\hline
p+Au~ \cite{Bog80}  &       7.5 GeV/c    & 50.$\pm$2.   &  33.$\pm$2.   \\
\hline
$^{20}$Ne+U~\cite{gutbrod} &250 A$\cdot$MeV     & 36.6$\pm$0.3$^*$  & 31.9$\pm$4.2$^*$  \\
\hline
$^{20}$Ne+U~\cite{gutbrod} &400 A$\cdot$MeV     & 46.5$\pm$1.3$^*$  & 37.3$\pm$3.3$^*$\\
\hline
      &         &$\langle E_{kin}(^3He)\rangle$&$\langle E_{kin}(^4He)\rangle$  \\
      &         &              (MeV)           &         (MeV)                   \\ 
\hline
p+Ag~\cite{Vol74}        &   1 GeV        & 17.7$\pm$0.3  & 12.7$\pm$0.2  \\
\hline
$^3$He+Ag~ \cite{Vio98}   &   55 MeV       & 19.8          & 15.6         \\
\hline
Ar+ Ni~\cite{Bor96}      &50-100 A$\cdot$MeV    &   32.5        &  26.2        \\
\hline
Xe+Sn~\cite{Bou99}      & 50 A$\cdot$MeV       &    51.0       &  33.0         \\
\hline
Ar+Xe~\cite{Moh96}       & 400 A$\cdot$MeV      & 139.6$\pm$2.5 &134.0$\pm$6.1  \\
\hline
Au+Au~\cite{Pog93}      & 100 A$\cdot$MeV      & 85.0$\pm$4.0  & 59.0$\pm$2.0  \\   
\hline
Au+Au~\cite{Pog93}      & 150 A$\cdot$MeV      & 101.0$\pm$4.0 & 77.0$\pm$5.0  \\
\hline     
Au+Au~\cite{Pog93}      & 250 A$\cdot$MeV      & 147.0$\pm$4.0 &136.0$\pm$3.0  \\
\noalign{\smallskip}\hline
\end{tabular}
\end{center}
\end{table}
 
An appropriate way to describe processes involving 
many particles is the statistical approach. The system characterized
in the initial stage by nonequilibrium distribution 
functions evolves towards equilibration as a result of interaction between 
particles. In this process the system runs through different states.
The first one can be considered as 
equilibration of the one-particle degrees of freedom. 
In the following the evolution toward total thermalization can be 
considered as progressive involving of higher order particle correlations. 
For finite expanding systems one cannot predict
in advance what kind of equilibration should be considered.

It is usually accepted that Intermediate Mass Fragments (IMF) observed
in multifragmentation processes are mainly produced 
in a state close to thermalization. This assumption is supported by the success 
of statistical multifragmentation models, e.g. SMM \cite{bondorf} and 
MMMC \cite{gross}. However, this conclusion may not be true for composite
particles like $^3$He and $^4$He which can be formed in earlier stages of 
the reaction. We guess that their unexpected behaviour results from an interplay
of different production mechanisms. 

\section{Production of composite particles by coalescence}
 
First we point at an alternative way to form clusters. 
We start from 
a distribution of nucleons in the phase space at some "freeze-out" time 
obtained after a dynamical evolution. Generally, non-uniform distributions
are conceivable. But in some experiments, e.g. central nucleus--nucleus 
collisions \cite{Rei97}, it is possible to select events which are
nearly isotropic in space and look like thermal ones. 
Therefore, in such cases, we can simply assume that the nucleons populate the
available {\em many-body} phase space uniformly, i.e. there is equilibration in 
one-particle degrees of freedom, that gives rise to a thermal distribution 
for individual nucleons in thermodynamical limit. 
 
A composite particle can be formed from two or more
nucleons if they are close to each other in 
the phase space. This simple prescription is known as coalescence model and it
reflects the properties of the nucleon--nucleon interaction. Here we use the 
coalescence in momentum space only. The basic assumption is that the dynamical 
process which leads to a momentum redistribution is very fast (nearly 
instantaneous) so that the coordinates of nucleons are just defined by their 
momenta. It is also justified taking into account quantum properties 
of the system since the wave functions of nucleons are rather broad.
This type of coalescence model has proven successful in reproducing 
experimental data (see e.g. \cite{BU63,Ton83,CS86}).

In the standard formulation of the model it is assumed that the fragment density in 
momentum space is proportional to the nucleon density times the probability of 
finding nucleons within a small sphere of the coalescence radius $p_0$. From  
this hypothesis an {\em analytical} expression can be derived
for differential yields of coalescent clusters\\ \cite{BU63,CS86}:
$$ 
\left( E_A \frac{d^3N_{A}}{d\bar{p}^{3}_{A}}\right) =
\frac{2S_A +1}{2^A} \frac{1}{\nu!}\frac{1}{z!}\left( \frac{4\pi}{3}
\frac{p_{0}^3}{m_n} \right)^{A-1} \left(
E_n \frac{d^3 N_{n}}{d^3 \bar{p}_{n}} \right)^{A} \quad 
$$
where $E_A ,E_n ,\bar{p}_A$ and $\bar{p}_n $ are the energies and the momenta 
of fragments and nucleons, $\nu$ and $z$ are the numbers of neutrons and 
protons ($A=\nu+z$); $S_A$ is the spin of the fragment with mass number $A$ and
$m_n \approx 0.94$\,GeV is the nucleon mass.
However, this equation disregards correlations between different clusters since 
the conservation of the nucleon number is not taken into account. 
Therefore, the above formulae is valid only 
for $N_{A=1}\gg  N_{A=2}\gg N_{A=3}$... In many reactions this condition is not
fulfilled. 

We developed an alternative formulation of the coalescence model which is
suitable for computer simulations. Nucleons can produce a cluster 
with mass number $A$ if their momenta relative 
to the center-of-mass moment of the cluster is less than $p_0$. 
Accordingly we take
$|\vec{p}_{i}-\vec{p}_{cm}|<p_{0}$ for all $i=1,...,A$, where 
$\vec{p}_{cm}=\frac{1}{A}\sum_{i=1}^{A}\vec{p}_{i}$. 
This is performed by comparing the momenta of all nucleons. 
In the following examples the value $p_0\approx94$\,MeV/c 
has been adopted corresponding to relative
velocities $v_{rel}$=0.1$c$ in agreement with previous analyses \cite{Ton83}. 

We note a problem which is sometimes disregarded in these simulations. 
Some nucleons may have such momenta that they can belong to 
different coalescent clusters according to the coalescence criterion. 
In these cases the final decision depends on the sequence of nucleons within
the algorithm. To avoid this uncertainty we developed 
an iterative coalescence procedure. $M$ steps are calculated in the coalescence 
routine with the radius $p_{0j}$ which is increased at each step $j$: 
$p_{0j}=(j/M)\cdot p_0$ ($j=1,...,M$). Clusters produced at earlier steps 
participate as a whole in the following steps. In this case the final clusters 
not only meet the coalescence criterion but also the nucleons have the minimum 
distance in the momentum space. Mathematically exact, this 
procedure gives correct results in the limit $M\rightarrow\infty$ but we found 
that in practical calculations it is sufficient to confine the steps to $M$=\,5. 

To demonstrate how fragments are produced by coalescence we take as an 
example a nuclear system with mass number $A_0=200$ and charge $Z_0=80$ 
at various excitation energies $E^*$=10, 20, 30 and 100 A$\cdot$MeV.
We disintegrate the system 
into nucleons by taking away about 7~A$\cdot$MeV (binding energy). The rest of the 
energy turns into the kinetic energies of nucleons which populate the whole 
available many-body momentum phase space uniformly. 
We use the procedure developed in \cite{kopylov} to generate momenta.  
\begin{figure}
\begin{center}
\epsfig{file=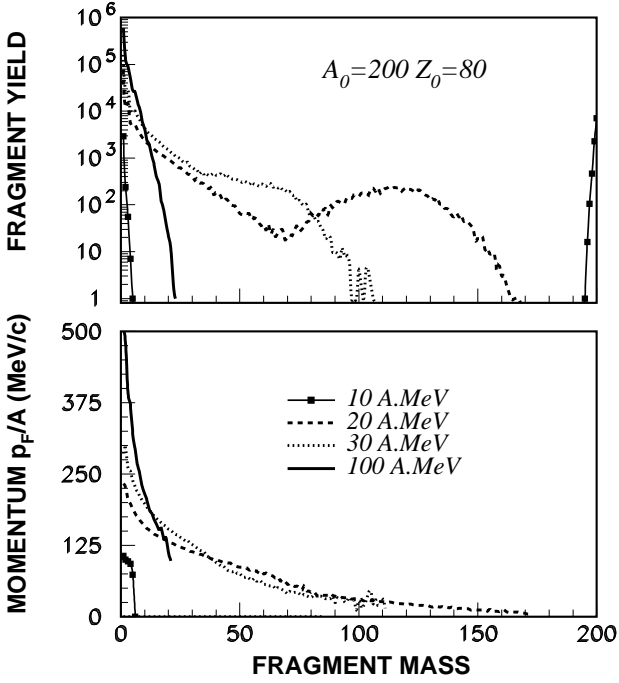,width=8.6cm}
\end{center}
\caption{Fragment production by coalescence. Upper panel:
         fragment yields per 10$^4$ generated events at 4
         different excitation energies. Lower panel: the
         corresponding fragment momenta per nucleon.}
\label{fig:simulation}
\end{figure}
Example mass distributions of generated fragments are shown in fig. 
\ref{fig:simulation}. At low excitation energies we see 
one big "coalescent" cluster and some light fragments (an "U--shape" distribution).
 With increasing excitation energy
the large cluster will be destroyed and several small clusters are produced. 
At about 30 A$\cdot$MeV a transition appears which changes the mass distribution
into a "step"-like one. At higher energies an exponential-like shape of the 
mass distribution occurs and that corresponds to the standard coalescence picture at high 
energies. We found that this 
transition happens at $\langle p \rangle / p_0 \approx 2$,
where $\langle p \rangle$ is the average momentum of thermalized nucleons, and 
it depends weakly on the size and other parameters of the system. 
In fig. \ref{fig:simulation} we show also the average 
momenta of the clusters. We see that the coalescence 
provides a special ordering in the kinetic energy of the fragments: light  
clusters have larger energies per nucleon than the heavy ones. 
Large clusters are preferentially formed 
from nucleons with small momenta, i.e. which are closer to the center of the 
phase space. 

The evolution of the mass distribution with excitation energy predicted by 
the coalescence model corresponds 
to our expectations for the decay of finite nuclei. Large clusters must be 
excited if they are treated in the classical coalescence model because there 
is a motion of nucleons with respect to the center of mass of the clusters.
In this case 
a big coalescent cluster can be treated as an excited remnant of the system 
in which more complete equilibration is achieved and which can decay in 
a thermal-like way. 
Such a picture is supported by dynamical calculations which show the different 
reaction stages in nucleus--nucleus collisions at higher energies 
(e.g. \cite{bond94}). After the first interactions
some nucleons immediately leave the system and we can assume that later on 
they form coalescent clusters in final state interactions. This fast emission 
lowers essentially the energy of the remaining part and favours its 
equilibration, so we have a chance to treat it by statistical methods.

\section{Kinetic energies of $^3$He and $^4$He}
 
\begin{figure}
\begin{center}
\epsfig{file=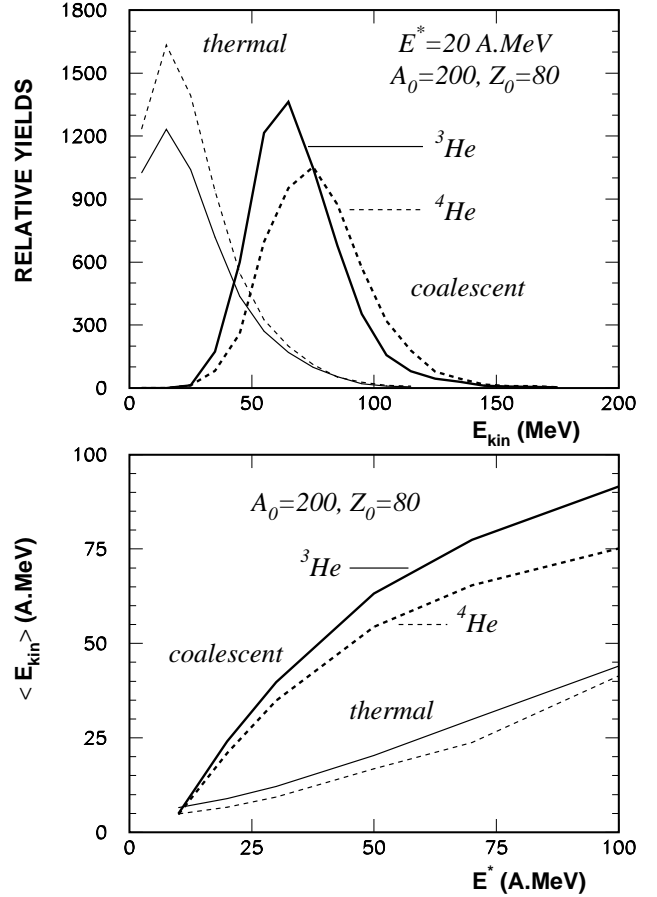,width=8.6cm}
\end{center}
\caption{Kinetic energy spectra (top) and mean kinetic energies per nucleon 
         (versus excitation energy, bottom) of $^3$He and $^4$He
         calculated for the thermal and coalescence
         scenaria.}   
\label{fig:spectra}
\end{figure}
Now we show qualitatively how the coalescence can explain the "puzzle".
In fig. \ref{fig:spectra} (top) we present the kinetic energy distributions of
coalescent clusters
$^3$He and $^4$He generated for $E^*=20\,$A$\cdot$MeV. The peaks in both spectra
are shifted towards higher energies similar to collective motion 
(radial flow). For a different (e.g. non-equilibrium) initial 
distribution of the nucleons in momentum space we expect broader spectra 
with possible shifts of the peaks. But, in any case, 
the kinetic energies of the clusters will be high. 
In the bottom of fig.~2 we plot the mean kinetic energies of helium isotopes 
for two scenaria and different excitation energies. 
The coalescent $^3$He clusters 
have slightly larger kinetic energy per nucleon compared with $^4$He. 
This energy difference increases with excitation energy though the total kinetic 
energy of $^3$He is still lower than the energy of $^4$He.   
This is caused by the mentioned correlation between the cluster production and 
the momenta of nucleons which form the clusters (fig.~1, bottom). 
Such a correlation was already under discussion, e.g. \cite{kunde},
to explain the observed  radial-flow energies of the fragments. 
In fig.~2 we also show the kinetic energies and 
spectra of $^3$He and $^4$He calculated with the 
SMM code \cite{bondorf} assuming a completely thermalized system at the same 
excitation energies. Herewith all thermal processes including deexcitation
of hot primary fragments are taken into account. There is a striking
difference in the kinetics of light clusters produced by  
coalescence or thermal mechanisms. 
In high energy reactions the difference may even increase 
because the energy released in the preequilibrium stage
is usually much larger than the remaining 
equilibrium energy. We have checked also the influence of secondary 
deexcitations of possibly excited intermediate coalescent clusters 
(with $A>4$) on the spectra of the helium isotopes: the energy decrease 
is relatively small and the difference to the thermal spectra remains 
pronounced. 

We expect that the thermal fragment production and the coalescence
coexists in  heavy--ion collisions over a wide energy range. The final
shapes of the fragment spectra depend on their relative contributions.
It is well known that the equilibrium decay provides a large number of
$\alpha$--particles, whereas 
$^3$He clusters are suppressed. But $^3$He clusters can be produced
predominantly by coalescence from the large number of emitted preequilibrium
nucleons. As a result, the kinetic energies of $^3$He become larger 
compared with those of the $\alpha$--clusters. Such conditions may
occur at intermediate energies in heavy ion reactions. However,
in central nucleus-nucleus collisions the size of the thermalized
source decreases with increasing beam energy
as shown by SMM calculations and comparison with experimental data 
\cite{Neu96,williams}. Thus at sufficient high energies the coalescence becomes a 
dominant production mechanism for both kinds of He clusters and their
kinetic energies may be very similar. 

\section{Comparison of experimental data with SMM and coalescence calculations}

We checked this hypothesis by calculations of fragment production 
in central $Au+Au$ collisions using the SMM model which was modified
by inclusion of radial flow and preequilibrium particle emission.
Recently this approach was used in ref's.\cite{bond94,Neu96,williams,dagostino}.
In the framework of this model it is assumed
that the total c.m. energy is shared between a thermal source 
(responsible for IMF and thermal LPC production), creation of
pions (taken phenomenologically from experimental data) and 
preequilibrium emission of nucleons. Afterwards, 
these nucleons may coalesce into composite particles.
A uniform initial distribution of "preequilibrium" nucleons in many-body 
momentum space is assumed. As already mentioned this assumption allows  
to estimate simply the influence of coalescence in the particular case of 
central events.

The excitation energy and mass 
of the equilibrium sources were parametrized by fitting the charge
distributions and multiplicities of IMF's 
in the beam energy range from 150 to 1050 A$\cdot$MeV 
as described in ref. \cite{Neu96}. 
The radial flow was found by fitting the kinetic energy spectra of IMF's
and it matches the experimental systematics \cite{Rei97,Rei98}. 
The source parameters at 1150 A$\cdot$MeV \cite{Lis95} obtained by extrapolation are  
very close to that at 1050 A$\cdot$MeV. 
In ref's. \cite{Neu96,williams} it has been concluded that the size of the thermal 
source decreases considerably with the beam energy, and preequilibrium 
emission becomes important. 
The part of energy attributed to the preequilibrium stage is very large and, 
therefore, the masses of the coalescence fragments decrease nearly 
exponentially. 
\begin{figure}
\begin{center}
\epsfig{file=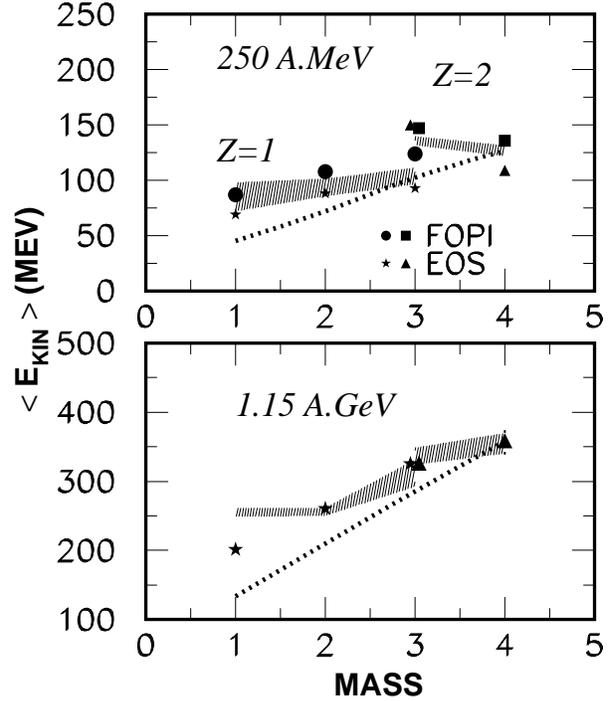,width=8.6cm}
\end{center}
\caption{Mean kinetic energies vs. mass number of LCP's in  
         central collisions of Au on Au  
         at incident energies 250 and 1150 A$\cdot$MeV.
         Black symbols: data, dotted
         line: thermal production only, 
         shadowed area: thermal orign and coalescence ($v_{rel}$=0.1$c$) 
         assuming an overall uncertainty of
         \mbox{$\simeq$ 15 \%} in the model parameters.} 
\label{fig:puzzle}
\end{figure} 
In fig. \ref{fig:puzzle} we compare the experimental data from central Au+Au
collisions with the calculated mean kinetic energies of light charged particles. 
The energy spectra of $^3$He and $^4$He were simulated for
250 A$\cdot$MeV in the angular range $60^0 \leq \Theta_{cm} \leq 90^0$ 
using the lower and upper registration thresholds given in ref.\cite{Pog93}.
LCP's emitted from the thermal source with superimposed collective motion 
show the expected increase of $\langle E_{kin} \rangle$ with $A$. However, 
the energy stored in preequilibrium nucleons is decisive for clusters 
predominantly produced from these nucleons: 
the $^3$He-$^4$He anomaly appears 
if a substantial part of $^3$He comes from coalescence whereas the thermal 
production of $^4$He dominates. 
In addition we show in \mbox{table 2} a reasonable reproduction of the hydrogen and 
helium multiplicities within this scenario.

\begin{figure}
\begin{center}
\epsfig{file=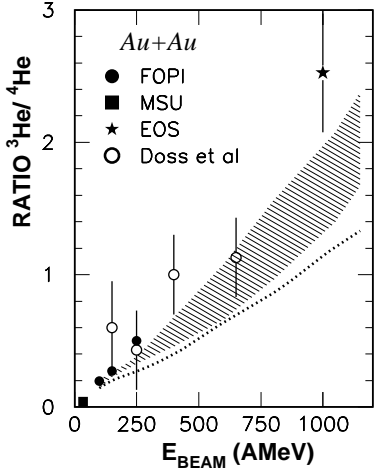,width=6.6cm}
\end{center}
\caption{Yield ratio $^3$He/$^4$He in central Au+Au collisions
         vs. beam energy. Symbols show the data, dotted line: equilibrium-source
         calculations with the
         parameter set from \cite{Neu96}, shadowed area:
         calculations including preequilibrium and coalescence,
         similar to fig.~3.}
\label{fig:ratio}
\end{figure}

Enhanced production of $^3$He with increasing beam energy has been established
in central Au+Au collisions \cite{Pog93,Lis96,Dos88} as shown in fig.
\ref{fig:ratio}. The $^3$He yield dominates over $^4$He production in the region
$\geq$ 600 A$\cdot$MeV. Calculations with our model confirm this trend. 
There is an obvious surplus of $^3$He by coalescence  
with respect to the thermal production. 
This behaviour seems to be related to the disappearance of the anomaly 
established at 1.15 A$\cdot$GeV \cite{Lis95}, fig.~\ref{fig:puzzle}, that is 
also present in our calculations. The disagreement in the calculated proton energies  
may be caused by the centrality criterion in ref.\cite{Lis95} which is
different from that in ref. \cite{Pog93}.
Selection of protons around $\Theta_{cm}\simeq$ 90$^o$ 
in events with high multiplicity may push down their kinetic energies since 
such a criterion includes many events with predominantly forward-backward
directed high energy nucleons. Indeed, lower proton energies were obtained
by simulations assuming anisotropic preequilibrium emission of nucleons whereas
the conclusion about the $^3$He-$^4$He-anomaly is maintained.

\section{Conclusions}

Summarizing we found that the appearance and disappearance of the 
$^3$He- $^4$He anomaly reflects an interplay of equilibrium and nonequilibrium 
processes in these reactions. Besides the thermal emission 
the coalescence of nonequilibrium nucleons 
was found to be the important process
which is responsible for enhanced mean kinetic energies of $^3$He. 
The effect is pronounced when both the thermal production and the coalescence 
contributions are comparable. At sufficient high energies 
only a small part of the colliding nuclei attains equilibration and  
both $^4$He and $^3$He are mainly produced by
coalescence. As a consequence, the anomalous behaviour vanishes.\\  
We mention another approach \cite{Bou99} which shows that 
in central nucleus-nucleus collisions around the Fermi energy a fast fragment 
emission during the expansion stage (described by the EES model \cite{Fri90}) 
can be responsible for the "puzzle". 
This explanation is qualitatively in agreement with our findings. However, we 
suppose that coalescence is a general mechanism which may be applied successfully 
in statistical or dynamical models (see e.g. \cite{gutbrod,Ono99,hombach}).
It seems that the correlation between energies and masses of the clusters
predicted by the coalescence model may explain the observed features of fragment
production in a wide energy range.\\
\begin{table}
\caption{Multiplicities of LCP's in central Au+Au collisions.
Typical sum uncertainties of the calculations are $\pm$ 5.}
\label{tab:mult}
\begin{center}
\begin{tabular}{|c|c|c|c|c|c|}
\hline
 &E(A$\cdot$MeV)&pre&equ&sum&data,\,ref.\cite{Rei97}\\
\hline
{\em Z}=1 &150     &32.2  &32.4  & 64.6       &61.84(0.58)\\ 
\hline
{\em Z}=1 &250     &52.2  &30.8  & 83.0       &75.82(0.62)\\ 
\hline
{\em Z}=1 &400     &60.0  &34.8  & 94.8       &92.04(0.62)\\
\hline\hline
{\em Z}=2 &150     &5.3   &15.6  & 20.9       &26.76(0.36)\\
\hline
{\em Z}=2 &250     &8.9   &14.5  & 23.4       &27.27(0.36)\\
\hline
{\em Z}=2 &400     &9.3   &14.3  & 23.6       &24.16(0.30)\\
\hline
\end{tabular}
\end{center}
\end{table}

This work was supported by the German Ministry of Education and 
Research (BMBF) under contract 06-DR-828. A.S.B. thanks the INFN 
(Bologna section) for hospitality and support. We acknowledge 
stimulating discussions with H.-W.Barz and W.Reisdorf.

\end{document}